# Iron Nuclei in Ultra-High Energy Cosmic Rays Near the Earth

A. V. Uryson

Lebedev Physical Institute of the Russian Academy of Sciences, Moscow, 119991 Russia

e-mail: *uryson@sci.lebedev.ru*

**Abstract**. The propagation of cosmic iron nuclei with energies $\geq 10^{19}$ eV from their sources to the Galaxy is discussed assuming that cosmic rays at ultra-high energies are of extragalactic origin. In extragalactic space, cosmic nuclei interact with background emissions and inevitably fragment. An analysis is performed of the fraction of iron nuclei reaching the Earth and how its energy depends on the distance from cosmic ray sources. It is found that energies of iron nuclei can be used to determine constrain on the distances from their sources.

## Introduction

The sources of ultra-high energy (UHE) cosmic rays (CRs) with $E > 4 \times 10^{19}$ eV have yet to be reliably established. Both galactic and extragalactic objects where CR particles can be accelerated to such energies are discussed in literature.

In this work, we suggest that CRs are of extragalactic origin. More specifically, CRs are accelerated to UHEs in the vicinity of the supermassive black holes (SMBHs) found at the center of many galaxies, including the Milky Way. It is now generally accepted that SMBHs exist in the core of almost every galaxy [1].

CRs can be accelerated in jets [2, 3]: in accretion disks [4]; near the polar caps of SMBHs, where particles fall from the accretion disk [5, 6]; and CRs can extract energy in the ergosphere of spinning black holes [7, 8].

If the hypothesis about particle acceleration in the vicinity of SMBHs is correct, then almost every galactic nucleus with an SMBH can be a source of CRs. (An active SMBH phase is required when the accretion disk is being fed if CRs are to be accelerated in a jet. Otherwise, the jet switches off.)

The jet and accretion disk of an SMBH contain stellar matter, which is why CRs at their sources include elements with a variety of mass numbers *A* up to iron nuclei. The elemental composition of CRs changes on their journey from the source, since the nuclei of UHE CRs interact with background radiation (the cosmic microwave background, radio waves, extragalactic light) in reactions:

$A + \gamma \rightarrow A + e^+ + e^-$ (direct pair production with the threshold energy in the center-of-mass system is 1 MeV);

$A + \gamma \rightarrow A' + mN + n\pi$ (photopion production with the threshold energy in the center-of-mass system is 145 MeV);

$A + \gamma \rightarrow A' + mN$ (the photodisintegration of nuclei with the threshold energy is the binding energy of nucleons in the nucleus, with the cross section of the process peaking at energies of tens of MeV).

The elemental composition of UHE CRs is currently a subject of intense research. The Telescope Array and Pierre Auger observatories have shown that CRs with energies of $\sim 10^{18}$ to $10^{20}$ eV contain nuclei of elements ranging from He to Fe [9, 10]. (The average composition of CRs in the range of energies from $10^{15}$ to $10^{18}$ eV is obtained at the Tunka-133 and TAIGA-HISCORE detectors [11].)

Photodisintegration of UHE CRs in extragalactic space was previously discussed in [12].

The propagation of different CR elements from the source to the observer was analyzed in [13, 14]. In this work, we discuss the propagation only of iron nuclei in intergalactic space. Interacting with background radiation iron nuclei fragment and particles with mass numbers $A = 1-55$ form on the way to an observatory. In this work, we calculate fractions $R$ of nuclei with mass numbers $A = 56$ (iron) and very close values $A = 54, 55$ (stable isotopes of iron and Mn) reached from the source to the observatory, relative to all other particles with $A = 1$ to $A = 56$:

$R = $(particles with $A = 54\text{-}56$) / (all particles with $A = 1\text{–}56$) . (1)

Calculations were made with the TransportCR code [15], that is publicly available on the Internet. In this code, photonuclear interactions are described according to [16].

**Model**

The main assumptions of the model concern the sources of UHE CRs and the interaction between particles and background radiation. They are the following.

I. Assumptions about the sources of CRs

1) Sources of UHE CRs are SMBHs in galactic nuclei.

2) Distances to sources are $L \approx 2\text{–}3500$ Mpc.

3) At such distances, we must consider the evolution of sources. The evolution of SMBHs is unclear, and the evolution of one type of active galactic nuclei - BL Lac - is considered in the calculations.

4) The injection spectrum in the sources is power-law $E^{-\gamma}$, where $\gamma = 2.2$.

5) UHE CRs are iron nuclei ($A = 56$). We assume this because we are interested in the propagation of iron nuclei only. Of course, both protons and nuclei of other elements are accelerated in the sources.

Items (3) and (4) are chosen because the bulk of data on CRs is described with them [17].

## II. Background radiation

1) The cosmic microwave background (CMB) has the Planck energy distribution. Average photon energy is $\varepsilon_r = 2.3 \times 10^{-4}$ eV, average density is density $n_r = 400$ cm$^{-3}$.

2) Background radio emission has the characteristics obtained in [18, 19].

3) The extragalactic background light (EBL) has the characteristics described in [20].

## Results

Energy dependence $R(E)$ in the range of energies ($3 \times 10^{18}$ to $3 \times 10^{20}$) eV for several values of redshift $z$ of sources is shown in Fig. 1. The figure shows that nuclei of $^{56}$Fe, including fragments of them with mass numbers $A = 54$ and 55 can arrive at an observatory both from nearby (within a radius of 40 Mpc) and remote sources (at distances up to 1000 Mpc). However, the fraction and the energy of the nuclei that reach the Earth are different depending on the distance to the source.

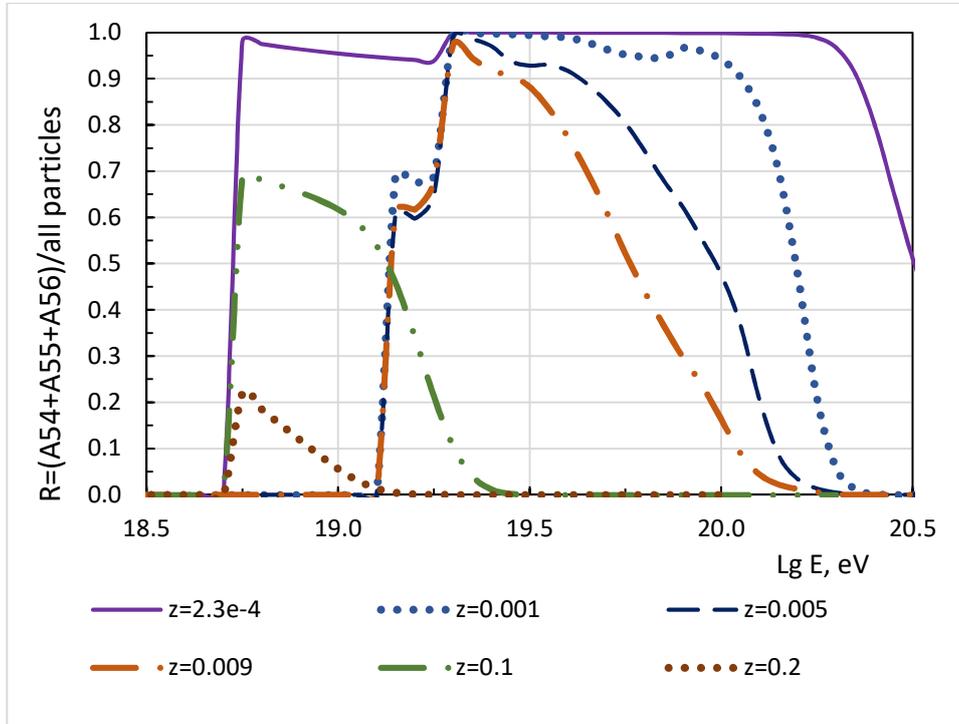

**Fig. 1.** The fraction $R$ of nuclei with $A = 54, 55, 56$ that reach the Earth from sources at various redshifts, as a function of the energy.

When sources are at red shifts $z < \sim 0.009$, the fraction of iron nuclei is $R \approx 1$ in the energy range of $\approx (5.6 \times 10^{18} - 2 \times 10^{20})$ eV. The shorter the distance to the source, the wider this interval is. The energy range where $R \approx 1$ decreases with increasing distances and shifts to lower energies. This is illustrated in Fig. 1 by curves calculated for sources with redshifts $z = 0.1$ and 0.2. The fraction of iron nuclei coming from such distances (i.e., hundreds of Mpc) is $R < 1$ at all particle energies. The maximum values for these nuclei are $R \approx 0.7$ and 0.2, respectively, at energies of $\approx 5.6 \times 10^{18}$ eV. Iron nuclei that came from distances $L > 200$ Mpc therefore do not fall into the range of ultra-high energies $E > 4 \times 10^{19}$ eV.

## Conclusion

Iron nuclei including their fragments with very close mass numbers $A = 54$ and 55, can arrive at an observatory from distances of up to several thousand Mpc. However, iron nuclei that came from distances $L > 200$ Mpc do not fall into the range of ultra-high energies.

In our model, the sources of UHE CRs are SMBHs in galactic nuclei. According to modern concepts, SMBHs exist in the cores of virtually all galaxies. Sources of $^{56}$Fe nuclei with energies $E > 10^{20}$ eV are therefore galaxies of the Local Group (the Magellanic Clouds and Andromeda (M31)). Sources of $^{56}$Fe nuclei with $E \approx 10^{20}$ eV are galaxies of the Local Supercluster, the one closest to the Local Group. Sources of $^{56}$Fe nuclei with $E < 3 \times 10^{19}$ eV are galaxies outside the Local Supercluster, located at distances of up to $L < \approx 100$ Mpc, and sources of $^{56}$Fe nuclei with $E < 10^{19}$ eV are galaxies at distances $L \sim 200–3500$ Mpc. A huge number of sources are located at such distances, so iron nuclei fall at an observatory having energies $E < 10^{19}$ eV.

In our model, the anisotropy of UHE CRs could be due mainly to the inhomogeneity in the spatial distribution of galaxies, and/or the spatial distribution of SMBHs in which vicinity CRs are accelerated.